\documentclass[12pt, onecolumn, letterpaper]{article}
\usepackage[utf8]{inputenc}
\usepackage{graphicx}
\usepackage{scicite}
\usepackage{times}
\usepackage{amssymb}
\newenvironment{sciabstract}{%
\begin{quote} \bf}
{\end{quote}}

\newcommand{\lya}{Ly$\alpha$}
\newcommand{\Lya}{Lyman~$\alpha$}
\newcommand{\HI}{H\textsc{i}}
\newcommand{\farcs}{$.\!\!^{\prime\prime}$}

\newcommand{\aj}{Astron.~J.}                   
\newcommand{\apj}{Astrophys.~J.}                 
\newcommand{\apjl}{Astrophys.~J.}      
\newcommand{\apjs}{Astrophys.~J. Suppl. Ser.} 
\newcommand{\aap}{Astron.~Astrophys.}  
\newcommand{\aaps}{Astron.~Astrophys.~Suppl.~Ser.}

\newcommand{\mnras}{Mon.~Not.~R.~Astron.~Soc.} 
\newcommand{\pasa}{Publ.~Astron.~Soc.~Aus.} 
\newcommand{\pasp}{Publ.~Astron.~Soc.~Pac.} 
\newcommand{\procspie}{Proc.~SPIE}   

\title{\textbf{\sffamily Gravitational lensing reveals ionizing ultraviolet photons escaping from a distant galaxy}}

\author{%
T. Emil Rivera-Thorsen$^{1\dagger\ast}$,
Håkon Dahle$^{1}$,
John Chisholm$^{2,3}$,\\
Michael K. Florian$^{4}$,
Max Gronke$^{5}$,
Jane R. Rigby $^{4}$,
Michael D. Gladders$^{6,7}$,\\
Guillaume Mahler$^{8}$,
Keren Sharon$^{8}$,
Matthew Bayliss$^{9,10}$
\\
\\
\footnotesize{$^{1}$Institute of Theoretical Astrophysics, University of Oslo,
Postboks 1029,0315 Oslo, Norway}\\
\footnotesize{$^{2}$Observatoire de Genève, Université de Genève, 51 Ch.\ des Maillettes, 1290 Versoix, Switzerland}\\
\footnotesize{$^{3}$Department of Astronomy and Astrophysics, University of California, Santa Cruz, CA 95064}\\
\footnotesize{$^{4}$Observational Cosmology Lab, NASA Goddard Space Flight Center,} \\
\footnotesize{8800 Greenbelt Rd., Greenbelt, MD 20771, USA}\\
\footnotesize{$^{5}$Department of Physics, University of California, Santa Barbara, CA 93106, USA}\\
\footnotesize{$^{6}$Department of Astronomy and Astrophysics, University of Chicago, Chicago, IL 60637, USA}\\
\footnotesize{$^{7}$Kavli Institute for Cosmological Physics, University of Chicago, Chicago, IL 60637, USA}\\
\footnotesize{$^{8}$Department of Astronomy, University of Michigan,} \\
\footnotesize{1085 South University Avenue, Ann Arbor, MI 48109, USA}\\
\footnotesize{$^{9}$Massachusetts Institute of Technology-Kavli Center for Astrophysics and Space Research,}\\
\footnotesize{ 77 Massachusetts Avenue, Cambridge, MA, 02139, USA}\\
\footnotesize{$^{10}$Department of Physics, University of Cincinnati, Cincinnati, OH 45221, USA}\\
\\
\footnotesize{$^{\ast}$Corresponding author: \texttt{trive@astro.su.se}}\\
\footnotesize{$^{\dagger}$Current address: Stockholm University, Dept. of
Astronomy,}\\
\footnotesize{AlbaNova Universitetscentrum,  SE-106 91 Stockholm, Sweden}\\
}
\date{}
\usepackage[insection]{minitoc}
\begin{document}
\mtcsetdepth{secttoc}{3}
\mtcsettitle{secttoc}{}
\mtcsettitle{sectlof}{}
\dosecttoc
\dosectlof
\faketableofcontents
\fakelistoffigures


\maketitle

\begin{sciabstract}
%
%
%
%
%
%
During the epoch of reionisation, neutral gas in the early Universe was ionized
by hard ultraviolet radiation emitted by young stars in the first galaxies. To
do so, ionizing ultraviolet photons must escape from the host galaxy. 
We present \emph{Hubble Space Telescope} observations of the gravitationally
lensed galaxy PSZ1-ARC G311.6602--18.4624, revealing bright, multiply-imaged
ionizing photon escape from a compact star-forming region through a narrow
channel in an optically thick gas. The gravitational lensing magnification shows
how ionizing photons escape this galaxy, contributing to the re-ionisation of
the Universe.  
The multiple sight lines to the source probe absorption by intergalactic neutral
hydrogen on 
scales of no more than a few hundred, perhaps even less than ten, parsec.

\end{sciabstract}

Less than a billion years after the Big Bang, the Universe
went through the epoch of reionization (EoR)\cite{Fan2006},
in which Lyman continuum radiation (LyC, ultraviolet light with wavelengths
below 912 Ångström, capable of ionizing hydrogen) from the first galaxies
ionized the previously neutral hydrogen in the intergalactic medium (IGM). 
To escape into the IGM, 
the ultraviolet photons must have avoided absorption by neutral hydrogen within
the host galaxy.
After the Universe became ionized and transparent to ionizing wavelengths, we 
expect many galaxies would still continue to emit LyC photons.
However, only a few dozen such galaxies have 
been found, 
either in the local
Universe\cite{Leitet2013,Izotov2018b,Izotov2018a,Borthakur2014} or at
intermediate redshifts ($1 \lesssim z \lesssim 4$)
\cite{Vanzella2016,Vanzella2018,Shapley2016,Bian2017,Fletcher2018}, leaving 
much of the radiation necessary to reionize the Universe unaccounted for.

Escape of LyC photons from galaxies is made possible by radiative or mechanical
feedback from e.g.\ turbulence or young, hot stars, which can 
ionize most of the surrounding gas, or carve out channels through optically
thick neutral
gas\cite{Zackrisson2013,Herenz2017a,Sunburst2017,Chisholm2018a,Bik2018}.  These
escape scenarios can be distinguished by the spectral shape of the \Lya\
(\lya) emission feature, an atomic emission line arising from the transition
between the ground state and the first excited state in neutral hydrogen, at a
wavelength of 1216 Å. \lya\ scatters resonantly in  the same neutral
hydrogen that absorbs LyC, and the \lya\ line shape is sensitive to its
kinematics, geometry, and other properties~\cite{Sunburst2017}.
 Escape
through an optically thin medium results in a double-peaked \Lya\ profile with
narrow separation between the peaks\cite{Verhamme2015,Dijkstra2016,Izotov2018b}.
This is what is typically observed in conjunction with confirmed LyC escape,
except in a few ambiguous cases\cite{Izotov2018b,Vanzella2018}.

In a riddled, optically thick neutral medium,
both \lya\ and LyC may escape freely through empty lines of sight, with little
to no interaction with neutral hydrogen.
\lya\ then appears 
as a bright, narrow peak centered at the wavelength of the transition.
If the 
neutral medium contains enough passageways of low optical
depth, the majority of \Lya\ photons will escape after scattering a small number
of times, 
so the narrow central peak will dominate the line shape~\cite{Sunburst2017}.
%
If only few, narrow channels are present, 
theory predicts that the probability for a given \lya\ photon to escape in a few
scattering events declines, and the probability of trapping in
resonant scattering in the denser neutral medium rises~\cite{Sunburst2017}.  
This produces a characteristic, triple-peaked profile with a narrow, bright peak
at line center, superimposed onto the typical broader, double-peaked line of an
optically thick medium~\cite{Behrens2014,Sunburst2017}.

Previous work has observed such a triple-peaked Lyman-$\alpha$ profile in
PSZ1-ARC G311.6602--18.4624, hereafter nicknamed the Sunburst
Arc~\cite{Sunburst2017}, a gravitationally lensed galaxy at redshift
$z=2.37$\cite{Sunburst2017}. The Sunburst is a single galaxy, lensed into at
least 12 images by a massive foreground galaxy cluster at
$z=0.44$\cite{Dahle2016}. The galaxy is young, strongly star forming, and shows
no sign of an active nucleus (see fig.~S1). The Lyman-$\alpha$ profile indicated
that it was a prime candidate for strong LyC escape through a narrow channel
oriented towards Earth\cite{Sunburst2017}. Such a LyC channel should appear as a
multiply-imaged, compact source coincident with some of the brightest regions
seen in the extended, non-ionizing stellar continuum\cite{Dahle2016} when
observed using telescopes with sufficient resolution and ultraviolet
capabilities.

We observed the rest-frame Lyman continuum in the Sunburst Arc using the Wide
Field Camera 3 
(\textsc{WFC3}) on the \emph{Hubble Space Telescope} (HST) using the broad-band
filter F275W. The wavelength cut-off of this filter matches
the lowest-energy limit of the LyC at the redshift of the lensed galaxy, with
only 0.5\% of the total filter throughput having wavelengths longer than 912 Å.
We combined the F275W observations with previous observations using the
\emph{HST} Advanced Camera for Surveys (ACS) and the broad-band F814W filter.  
At the redshift of the lensed galaxy, F814W
is sensitive to non-ionizing near-UV light emitted mainly by the same
young, hot stars as LyC, but is not absorbed by neutral hydrogen.

\begin{figure*}[!ht] 
    \centering
        \includegraphics[width=.995\textwidth]{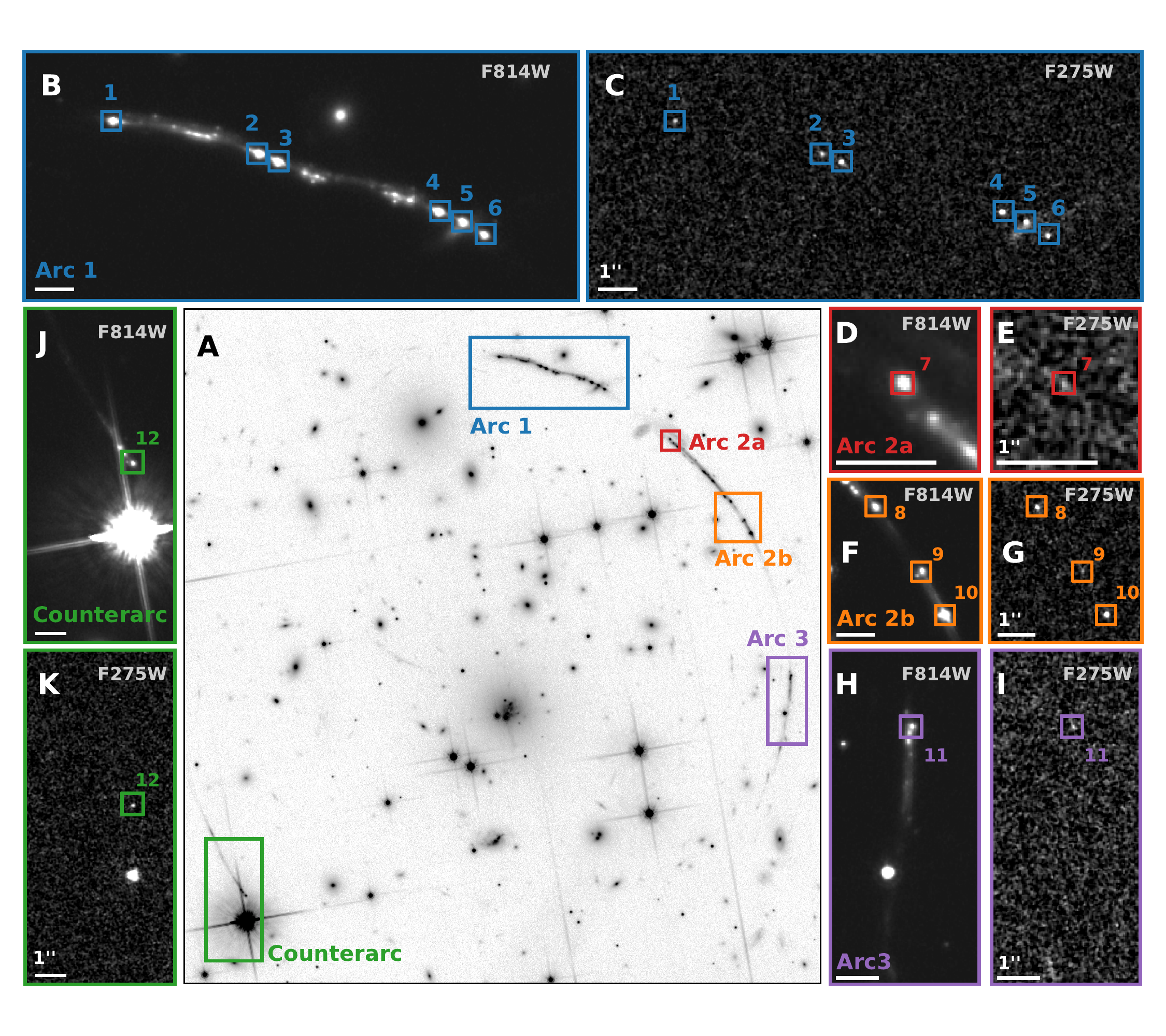}
	\caption{\textbf{Comparison of ionizing and non-ionizing morphology}.
		Panel A shows an overview of the entire arc, in the 
		\emph{Hubble} ACS F814W filter, with cutout locations marked.  The
		surrounding panels B-K show paired F814W and F275W observations, 
		zoomed in on the regions with confirmed Lyman Continuum 
		detection. All panels are oriented North up and East to the 
		left. Data in panels B-K are stretched by a
		hyperbolic arcsine function to balance visibility of faint
		features in F814W with noise in F275W panels. Panel A, in which
		noise is negligible, is scaled by a square root function to
		enhance faint features further, and colors are inverted to
		distinguish it from the zoomed in panels. All F814W panels
		except A share the same cut levels. All F275W panels share cut 
		levels, except E and I which due to the fainter sources are cut 
		off at 60\% the upper level of the rest.  The bright object near 
		image 12 is a foreground star.~\label{fig:cutouts}} 
\end{figure*}

Figure 1A shows the F814W image of the lens and arc system, while Fig. 1B-K show
close-ups of regions with emission in the F275W  filter. 
Each region is shown in both the ionizing and non-ionizing wavelengths, with the
images of the LyC emitting complex marked in both filters for comparison.
Image~5 is contaminated by UV continuum from a foreground galaxy which
contributes $\lesssim 5\%$ to its measured flux in F814W.  Lens modeling and
spectroscopy show that the 12 image-plane sources of ionizing LyC emission are
all lensed images of the same bright region [\cite{Methods}, figure S2], with
signal-to-noise ratios ranging from 4 to 42 (Table 1).  The physical diameter of
the LyC emitting region has an upper limit of $\sim 160$~pc\@\cite{Methods},
much smaller than the galaxy as a whole and consistent with star-forming regions
in local galaxies\cite{Adamo2013}. 

\begin{table*}[]
    \centering
    \begin{tabular}{lccccc}
        \hline
Image  & $m_{\mathrm{AB}}^{F275W}$ & S/N$_{\mathrm{\,F275W}}$
       & $m_{\mathrm{AB}}^{F814W}$ & $f_{\mathrm{esc,rel}} \times
T_{\mathrm{IGM}}$ & RA,DEC (hourangle, angle) \\
        \hline
1 &26.96$\pm$0.09& $12 $ & 22.91 & 23\% $\pm$ 4\% &15:50:07.29, -78:10:57.2\\
2 &26.84$\pm$0.08& $13 $ & 22.57 & 19\% $\pm$ 3\% &15:50:06.03, -78:10:58.1\\
3 &26.18$\pm$0.05& $23 $ & 22.68 & 40\% $\pm$ 5\% &15:50:05.86, -78:10:58.3\\
4 &25.75$\pm$0.03& $37 $ & 22.46 & 49\% $\pm$ 6\% &15:50:04.48, -78:10:59.6\\
5 &26.23$\pm$0.05& $23 $ & 23.08 & 55\% $\pm$ 7\% &15:50:04.27, -78:10:59.9\\
6 &26.43$\pm$0.06& $19 $ & 23.00 & 43\% $\pm$ 6\% &15:50:04.08, -78:11:00.3\\
7 &28.24$\pm$0.30& $3.7$ & 23.68 & 14\% $\pm$ 5\% &15:50:02.10, -78:11:04.9\\
8 &26.35$\pm$0.06& $19 $ & 23.18 & 54\% $\pm$ 7\% &15:50:00.25, -78:11:10.7\\
9 &27.34$\pm$0.15& $7.2$ & 23.44 & 27\% $\pm$ 5\% &15:49:59.84, -78:11:12.5\\
10&25.40$\pm$0.03& $42 $ & 22.35 & 61\% $\pm$ 8\% &15:49:59.63, -78:11:13.6\\
11&27.58$\pm$0.18& $6.2$ & 23.85 & 32\% $\pm$ 7\% &15:49:58.42, -78:11:26.9\\
12&26.65$\pm$0.08& $15 $ & 23.65 & 64\% $\pm$ 9\% &15:50:14.97, -78:11:47.5\\
        \hline
    \end{tabular}
    \caption{\textbf{Properties of regions with detected LyC escape.}
    The first column lists the lensed image numbers as designated in figure
    1. The second and third columns show the measured apparent magnitude and
    signal-to-noise ratio of the LyC (F275W) observations, and 
    the fourth column the apparent magnitudes in non-ionizing UV (F814W).
    The fifth column shows the apparent relative escape fractions (see main text)
    measured for each image. The corresponding apparent absolute escape
    fraction can be find by simply multiplying the relative escape fractions by
    a factor of $0.34\pm0.04$, which encodes the effect of internal dust in the
    galaxy~\cite{Methods}. The sixth and last column shows the
    celestial right ascencion and declination in hour:minute:second, 
    degree:minute:second.  Apparent magnitudes are corrected for Milky Way dust 
    reddening. Flux uncertainties in F814W are all $<$ 0.1\%.\label{tab:snr}}
\end{table*}

Measured magnitudes in both filters are tabulated in table~\ref{tab:snr}, along
with the computed apparent escape fraction. 
We computed ionizing escape fractions based on theoretical models of stellar
populations~\cite{Methods} which were fitted to 
spectroscopic observations of the non-ionizing wavelengths emitted
\cite{Sunburst2017,Methods} by the star forming region. From these model
spectra, we predicted the intrinsic flux ratios in the F275W and F814W filters,
and compared these to the observed ratios~\cite{Methods}. We derived both the
relative and absolute escape fractions, defined as the fraction of
dust-attenuated (relative) and total (absolute) ionizing radiation that
escapes the neutral gas in the galaxy. 

The observed flux in F275W is the radiation surviving absorption both within the
source galaxy and in the IGM\@. Consequently, the escape fraction we derive from
the flux ratios in the two filters is the combined effect of the internal and
intergalactic neutral hydrogen (\HI).  This quantity, found as the escape
fraction from the galaxy ($f_{\mathrm{esc}}$) times the transmission coefficient
of the IGM ($T_{\mathrm{IGM}}$), we refer to as the apparent escape fraction
$f_{\mathrm{esc}}^*$. The highest measured apparent escape fraction of
64$\pm$9\% (image 12) forms a lower limit to the escape fraction, corresponding
to the case of a completely transparent IGM\@.  Conversely, the apparent escape
fraction in image~12 provides a lower limit to the IGM transmission of
$T_{\mathrm{IGM}}\gtrsim 64\%$, as lower transmission coefficients would imply
an escape fraction higher than 100\%.  To further constrain the line-of-sight
escape fraction, we used the $T_{\textsc{igm}}$ distribution along simulated
lines of sight from a previous publication~\cite{Vasei2016}, and excluded the
values which would lead to an escape fraction larger than 100\%.  From the
remaining $T_{\mathrm{IGM}}$ distribution, we have extracted the 16th, 50th, and
84th percentiles and computed the corresponding escape fractions for
image~12~\cite{Methods}.  Figure~2 shows the apparent relative
($f_{\mathrm{esc,rel}}^*$) and absolute ($f_{\mathrm{esc,abs}}^*$) ionizing
escape fractions along our line of sight for each lensed image, with propagated
measurement uncertainties.
Absolute and relative escape fractions are shown for image~12 based on the IGM
transmission distribution discussed above, the 16th and 84th percentiles, and
the full range bracketed by an escape fraction of 100\% and the far upper end of
the $T_{\mathrm{IGM}}$ distribution~\cite{Methods}.  
region coding in fig.~1.  We estimate a relative line-of-sight ionizing escape
fraction of $f_{\mathrm{esc,rel}} = 93_{-11}^{+7}\%$, with $46\%$ as a robust
lower limit (assuming a completely transparent IGM and a flux ratio $2\sigma$
below the measured value). Correcting for modeled dust absorption internal to
the galaxy, we find the corresponding absolute line-of-sight escape fraction
$f_{\mathrm{esc,abs}} = 32^{+2}_{-4}\%$.
There is however not 
a simple correspondence between the line-of-sight escape fraction often reported
in the literature and the global ionizing escape fraction from a galaxy, unless
the galaxy is perfectly isotropic. 
The \lya\ profile of the Sunburst Arc~\cite{Sunburst2017} indicates that its
global escape fraction is much lower than the one reported here.

The computed escape fractions are subject to systematic uncertainties.  The
spectra used to construct the stellar populations are extracted in a larger
aperture than the ones used the photometric measurements, which could affect the
measured escape fractions~\cite{Methods}.
The F814W filter, which is used to photometrically calibrate
the stellar population model and predict the intrinsic LyC flux, covers an
adjacent but different wavelength range than the spectra to which the stellar
population model is fitted. Thus, we can not directly test the accuracy of the
model in the given range when comparing it to the photometric observations.
Finally, assumed stellar population models and dust attenuation laws introduce
systematic uncertainties. These systematic uncertainties can affect the inferred
escape fractions, but not the differences between them~\cite{Methods}.

\begin{figure}[ht]
    \centering
    \includegraphics[width=\columnwidth]{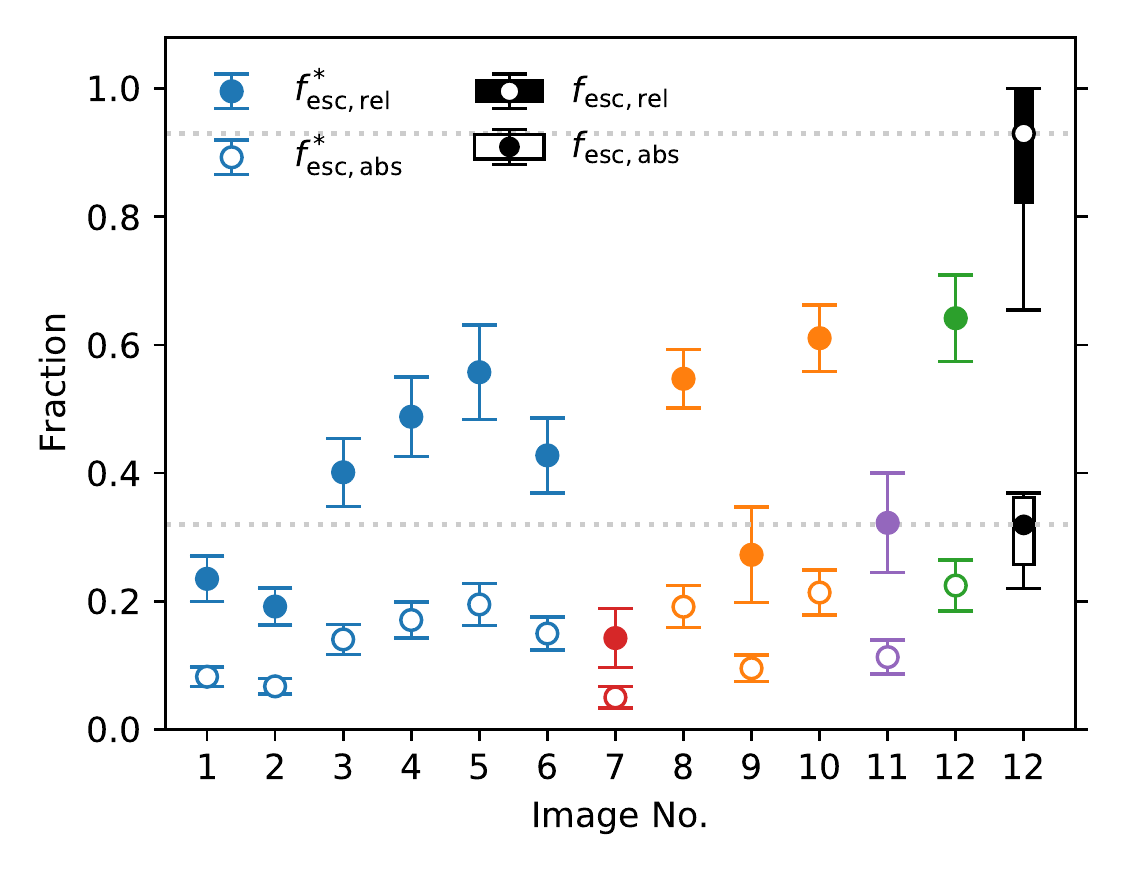}
    \caption{\small  \textbf{Lyman Continuum escape fractions}. For each 
		gravitationally lensed image, we show the fraction of the 
		dust-attenuated (relative; filled circles) and total 
		(absolute; empty circles) Lyman-continuum photons that reach 
		the telescope (apparent escape fraction, 
		$f^*_{\textnormal{esc}}$).  Colors correspond to those in fig.~1. 
		The black and white box-and-whiskers for image~12 show the 
		median value (dot), 16th to 84th percentile (box) and the full 
		allowed range (whiskers) of the relative (filled) and absolute 
		(outlined) escape fraction corrected for modeled IGM absorption.
		Values are listed in Table 1.\label{fig:fesc}}
\end{figure}

Lensing models [see fig.~S2 and~\cite{Methods}] show that all ionizing sources
in Arc 1, 3 and 4 (the Counterarc) are lensed images of the same system.  Arc 2
has complicated lensing geometry and has not yet been possible to model
completely,
but from the other arc segments, we find it likely that the ionizing
detections in Arc 2 are also images of the same system.  This is supported by
comparison of rest-frame near-UV 
spectra of 5 locations in Arcs~1 and 2, 4 that do and 1 that does not have
detected LyC. Comparison of their stellar C \textsc{iv} 1550 \AA\
and their interstellar and circumgalactic Si \textsc{iv}  1393,1402 \AA\
features show that the four ionizing locations are indistinguishable, while the
non-ionizing location is unambiguously different.
[\cite{Methods} and fig.~S2].

The lensing models indicate that the magnification factor in Arc 1 is between 10
and 30 for each image.  The central Lyman-continuum source is unresolved in all
images, which places an upper limit on the source size at the instrument point
spread function (PSF) of $0.09''$, corresponding to around 500 pc\ in the lens
plane.  For a magnification of 10 (in area), this corresponds to a physical
diameter of $\sim 160$ parsec in the source plane.  With a magnification of 30,
the maximum diameter would be $\sim 90$ pc, consistent with star forming regions
in local galaxies\cite{Adamo2013} and at higher
redshifts\cite{Adamo2013,Johnson2017}. 

\begin{figure}[ht]
    \centering
    \includegraphics[width=\columnwidth]{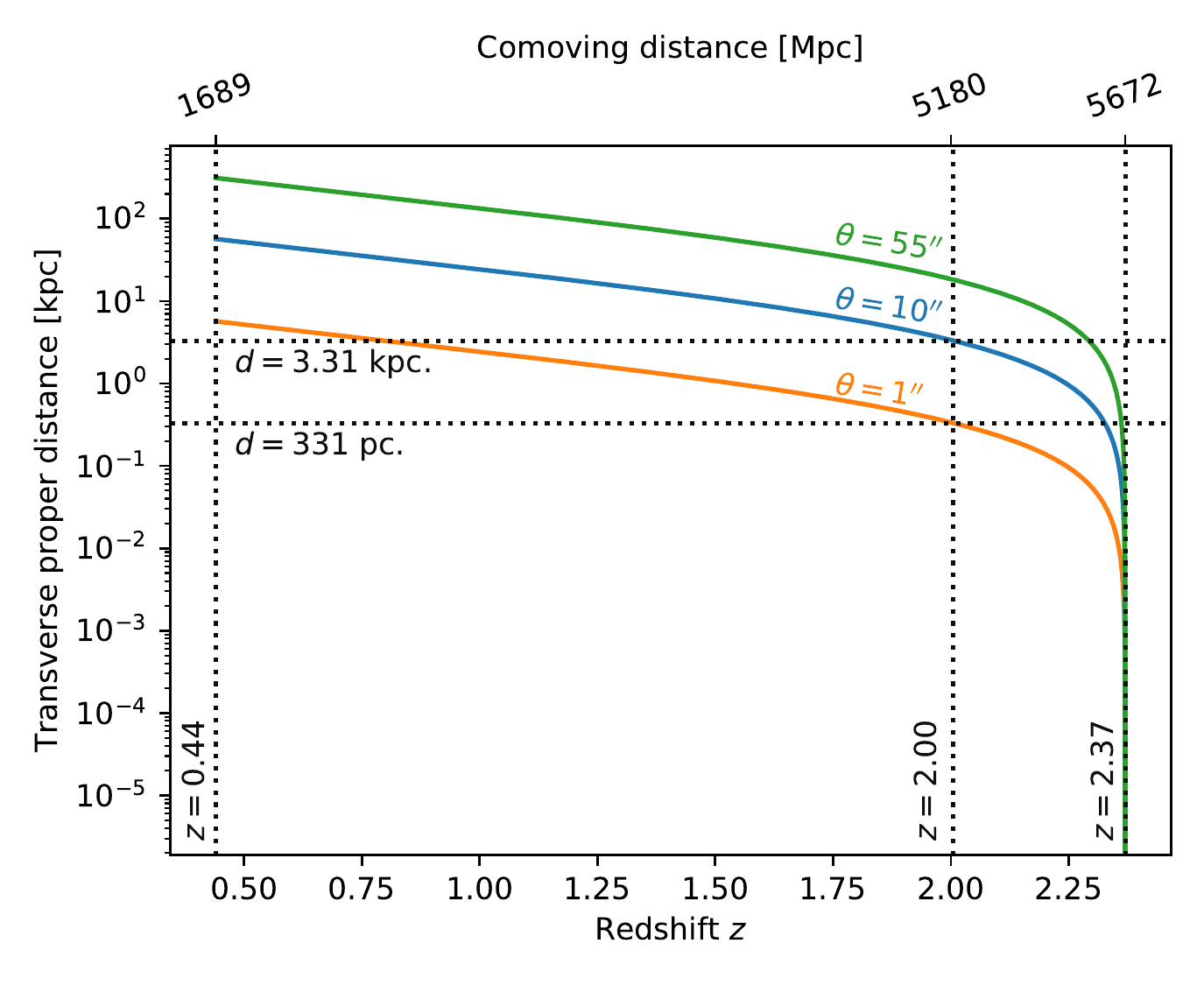}
    \caption{
        \small  \textbf{Transverse physical distances between lines of sight}. 
		Transverse physical separation $d$ of lines of sight with 
		angular separations $\theta$ of 1, 10 and 55$''$ (corresponding 
		to the distances between images~2--3, 1--6, and 6--12, 
		respectively) in the lensing plane are shown as a function of 
		redshift. The redshifts of the lensing plane ($z=0.44$), the 
		lensed galaxy ($z=2.37$), and the minimum redshift of 
		intervening H~\textsc{i} absorption ($z\lesssim2.0$) are marked, 
		along with the maximum transverse distances at the point of 
		H~\textsc{i} absorption.\label{fig:sightlines}}
\end{figure}

Light from the multiple lens-plane images of the source galaxy traverses
different paths between their point of origin and Earth. We interpret the
variations in $f^*_{\textnormal{esc}} $ as being due to varying amounts of
absorbing neutral Hyrdogen being present along these paths. Furthermore, the
absorption most likely occurs at $z \gtrsim 2$, at which redshift half of the
LyC emission detected in the F275W image has been redshifted beyond the
ionization limit of hydrogen~\cite{Methods}.
In fig.~3, we show the transverse distance between lines of sight towards
images~2 and 3 (one of the closest pairs), 1 and 6 (typical size of an arc
segment), and 1 and 12 (on opposite sides of the arc) as a function of redshift,
and mark the transverse distance between the lines of sight towards images~2 and
3 at $z=2$.  At this redshift, the physical transverse distance between the
lines of sight to images and 3 is $d \approx 330$ parsec. 

The upper limit on the redshift of absorption depends on the combined extent of
the ionizing source(s). For diameters of 160 or 30 parsec, the absorption must
occur at redshifts of $z \lesssim 2.27$ or $z \lesssim 2.35$, respectively.
Even for a very compact cluster a few parsecs across, the absorption must still
be $\gtrsim 1$ megaparsec (Mpc) outside the galaxy. Possible absorbers include
an undetected, interloping galaxy, or circumgalactic or intergalactic systems of
neutral hydrogen. 
It is also possible that the LyC originates in just one or a few very massive
O-type or Wolf-Rayet stars~\cite{Ramachandran2018}, in which case the absorption
could occur in the interstellar or circumgalactic medium (ISM or CGM,
respectively)  of the source galaxy itself, and the transverse separation of
sight lines could be a fraction of a parsec.

We considered differential magnification as a possible alternative explanation
of the differences in apparent escape fraction. However, this effect would lead
to a correlation between observed F814W magnitudes and $f^*_{\textnormal{esc}}$.
With a Pearson's $r = 0.2$ and a significance of $p = 0.53$, we do not find such
a correlation [Fig S5,~\cite{Methods}]. 


\subsubsection*{Supplementary materials}
\begin{itemize}
	\item[-] Materials and Methods 
	\item[-] Supplementary Text
	\item[-] Supplementary figures S1--S5
	\item[-] References (28--55)
\end{itemize}


\subsubsection*{Acknowledgements} E.R-T thanks Stockholm University for their
kind hospitality, Matthew Hayes and Angela Adamo for informative and helpful
conversations, and Kaveh Vasei and coauthors for generously sharing his
intermediate science products.

\subsubsection*{Funding}
E.R-T and H.D. acknowledge support from the Research Council of Norway.
M.G.\ was supported by by NASA through the NASA Hubble
Fellowship grant \#HST--HF2--51409 awarded by the Space Telescope Science
Institute, which is operated by the Association of Universities for
Research in Astronomy, Inc., for NASA, under contract NAS5--26555.

\subsubsection*{Author contributions}

E.R-T.\ led the project and paper writing, taking input and feedback from all
co-authors, especially H.D.\ and M.G. 
E.R-T.\ made all figures. H.D.\ wrote
the \emph{HST} proposals leading to the F275W and F814W observations, assisted
by E.R-T. M.K.F.\ reduced and combined the images in both filters. E.R-T.\
performed photometry and computed escape fractions. 
J.C.\ performed stellar population
synthesis  based on spectroscopic observations made by J.R.\ and
M.B and reduced by J.R. 
M.D.G.\ constructed the spatial model of flux in the MagE aperture.
K.S.\ and G.M.\ produced the lens model.

\subsubsection*{Competing interests}
The authors declare no competing interests.

\subsubsection*{Data and materials availability}
\begin{itemize}
	\item[-] Raw \emph{Hubble Space Telescope} WFC3 F275W and ACS
		F814W observations are
		available at the Mikulski Archive for Space Telescopes
		(MAST, \texttt{http://archive.stsci.edu/hst/}), program IDs
		15418 (F275W) and 15101 (F814W).
	\item[-] Raw Magellan/MagE observations and corresponding calibration data
		are available for download at Figshare~\cite{RigbyMageSunburst}
	\item[-] Parameter file to generate our lens model with
		Lenstool~\cite{LENSTOOL}, along with a DS9 region file showing
		the critical curves of this model, is available for download
		as online supplementary materials (Data S1). 
	\item[-] Scripts used to generate stellar population models are available
		for download at Zenodo~\cite{ChisholmScienceCode}.
\end{itemize}

\makeatletter
\renewcommand{\thefigure}{S\@arabic\c@figure}
\makeatother
\setcounter{figure}{0}

\makeatletter
\renewcommand{\theequation}{S\@arabic\c@equation}
\makeatother
\setcounter{equation}{0}

\clearpage
\newpage
\appendix
\setcounter{page}{1}

\begin{titlepage}
\begin{center}
    \vspace*{-1em}
    \includegraphics{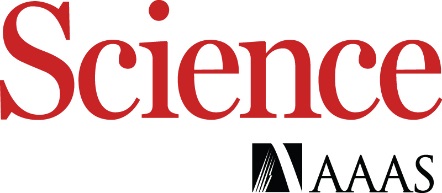} \\
    \vspace{3.em}
    \Large{Supplementary Materials for} \\
    \vspace{0.5em}
    \Large{\textbf{\sffamily Gravitational lensing reveals ionizing
    ultraviolet photons escaping from a distant galaxy}} \\
    \vspace{1em}
    \normalsize
    T. Emil Rivera-Thorsen$^{1\ast}$,
    Håkon Dahle$^{1}$,
    John Chisholm$^{2,3}$,\\
    Michael K. Florian$^{4}$,
    Max Gronke$^{5}$,
    Jane R. Rigby $^{4}$,
    Michael D. Gladders$^{6,7}$,\\
    Guillaume Mahler$^{8}$,
    Keren Sharon$^{8}$,
    Matthew Bayliss$^{9,10}$
    \\
    \vspace{0.8em}
    \footnotesize{$^{1}$Institute of Theoretical Astrophysics, University of
	    Oslo, Postboks 1029,0315 Oslo, Norway}\\
    \footnotesize{$^{2}$Observatoire de Genève, Université de Genève, 
            51 Ch.\ des Maillettes, 1290 Versoix, Switzerland}\\
    \footnotesize{$^{3}$Department of Astronomy and Astrophysics, University of
            California, Santa Cruz, CA 95064}\\
    \footnotesize{$^{4}$Observational Cosmology Lab, NASA Goddard Space Flight
            Center,} \\
    \footnotesize{8800 Greenbelt Rd., Greenbelt, MD 20771, USA}\\
    \footnotesize{$^{5}$Department of Physics, University of California, 
            Santa Barbara, CA 93106, USA}\\
    \footnotesize{$^{6}$Department of Astronomy and Astrophysics, University 
            of Chicago, Chicago, IL 60637, USA}\\
    \footnotesize{$^{7}$Kavli Institute for Cosmological Physics, University 
            of Chicago, Chicago, IL 60637, USA}\\
    \footnotesize{$^{8}$Department of Astronomy, University of Michigan,} \\
    \footnotesize{1085 South University Avenue, Ann Arbor, MI 48109, USA}\\
    \footnotesize{$^{9}$Massachusetts Institute of Technology-Kavli Center 
            for Astrophysics and Space Research,}\\
    \footnotesize{ 77 Massachusetts Avenue, Cambridge, MA, 02139, USA}\\
    \footnotesize{$^{10}$Department of Physics, University of Cincinnati, 
            Cincinnati, OH 45221, USA}\\
    \vspace{0.5em}
    $^{\ast}$\textbf{Correspondence to: \texttt{trive@astro.su.se}}\\
\end{center}
\vspace{2em}
\large{\textbf{This PDF file includes:}}
\begin{itemize}
	\setlength\itemsep{0em}
	\item[-] Materials and Methods 
	\item[-] Supplementary Text
	\item[-] Figures S1--S6
	\item[-] References 28--56
\end{itemize}
\end{titlepage}

\section{Supplementary materials}
\secttoc
\sectlof

\subsection{Materials and Methods}
\subsubsection{Conventions}

We have assumed a flat $\Lambda$-Cold Dark Matter cosmology with Hubble
expansion parameter $H_0 = 70$~km~s$^{-1}$~Mpc$^{-1}$ and  a matter density
parameter $\Omega_{M,0} = 0.3$.  Flux densities are given with respect to
wavelength ($f_{\lambda}$), and magnitudes are given in the $AB$ system.
Celestial coordinates are given in the ICRS system. The galaxy is designated
PSZ1-ARC~G311.6602--18.4624 as in the discovery paper~\cite{Dahle2016}.  The arc
segments referred to in the discovery paper~\cite{Dahle2016} as S1, S2 and S3
are here referred to as Arc 1, Arc 2 and Arc 3, respectively. The counterarc was
not yet confirmed as part of the lensed system in the discovery paper.

\subsubsection{\textit{Hubble Space Telescope} observations and data reduction}

The arc was observed in the UVIS channel of \textit{HST}
WFC3~\cite{WFC3Handbook} and \emph{HST} ACS~\cite{ACSHandbook} using the filters
F275W (program ID 15418, PI: H.  Dahle), and F814W (program ID 15101, PI: H.
Dahle).  The F275W observations, which capture Lyman continuum emission at the
redshift of the arc, were carried out during two visits, one on 2018 April 8,
and one on 2018 April 14 Universal Time (UT). The cumulative exposure time in
the F275W filter was 9422 s.  In the F814W filter, eight exposures were taken
for a total of 5280 s.\ between UT 2018 February 21 and UT 2018 February 22.
All observations were conducted using a 4-point dither pattern to minimize the
effects of bad pixels and to better sample the point spread function, improving
the effective resolution of the final data products.  The images in each filter
were aligned using the \textsc{drizzlepac}\cite{Gonzaga2012} routine 
\texttt{tweakreg}, and drizzled to a common grid with a pixel size of $0.03''$ 
with \textsc{astrodrizzle} using a Gaussian kernel and a ``drop size''
(\texttt{final\_pixfrac}) of 0.8.

\subsubsection{MagE observations and data reduction}

Spectra with the Magellan Echellette spectrograph [MagE,~\cite{MagE2008}] on the
Magellan Baade telescope were obtained from four observing runs in May 2017,
July 2017, April 2018, and August 2018, that covered 5 distinct image plane
regions in arc segments 1 and 2, see fig.~\ref{fig:magespecs} for slit
positions. Each region was observed with a total integration time of 2–4 hours.
The $1''$ slit was used for the first two runs, and the $0.85''$ slit was used
for the latter two runs, producing effective spectral resolving powers of  
$R \approx 4500$ and $R\approx 5200$, respectively, as measured from night sky
lines.  
The MagE data were reduced as described in~\cite{RigbyMagasauraI}, including
flat-fielding, wavelength calibration, sky subtraction, extraction, flux
calibration, telluric correction, and removal of Milky Way reddening.  The
results are one-dimensional flux- and wavelength-calibrated spectra covering
3200–8280 Å.

\subsubsection{Photometry\label{sec:photometry}}

We performed photometry using the source detection and photometry software
\textsc{Source Extractor}\cite{sextractor} running in dual mode using the F275W
observations as the detection image. The detection frames were smoothed by a narrow
kernel 1.5 pixels wide to avoid spurious detections due to single noisy pixels,
but fluxes were measured in the raw science frames in both filters. We extracted
the fluxes in a fixed aperture 4 pixels wide at the positions of the 12 images
in both of the filters.

Photometric uncertainties are dominated by the F275W observations. In F814W, the
signal-to-noise ratio is $> 1000$ in all photometric apertures.  To measure the
uncertainties in F275W, we placed 300 apertures at random positions containing
only background near the source, and computed the standard deviation of these
measurements. These were adopted as the standard errors for a given aperture,
and propagated forward in our analysis.  In a few cases, apertures placed on
nearby stars had to be rejected, as they artificially inflated the standard
deviations strongly.

The aperture size was selected to to optimize the balance between maximizing
signal-to-noise and robustness to aperture placement (which both favor larger
apertures), and minimizing contamination by the surrounding stellar population
which is detected in F814W only (favors smaller apertures).  While the Lyman
continuum emitting cluster complex is unresolved or barely resolved in the
\emph{HST} observations, the observations in F814W show a complex morphology of
clusters and underlying, diffuse stellar population. Thus, the measured
F275W/F814W colors will depend on the chosen aperture size: larger apertures
will include only the faint wings of the point spread function for the point
source images in F275W, while they will include a growing contribution from the
non-leaking stellar population in F814W. To determine the most appropriate
aperture size, we extracted fluxes and computed flux ratios for apertures sized
2, 4, 7, and 11 pixels, corresponding to approximately 1/2, 1, 2, 3, and 4 times
the full width at half maximum (FWHM) of the corrected PSF.\@ We found that for
apertures sizes $s \leq 4$ pixels, there were little difference between the
measured flux ratios, reflecting that the flux inside this aperture is dominated
by the leaking point sources. We thus opted for the 4 pixel aperture, to get the
best possible balance between larger aperture and uncontaminated flux from the
central region. We applied aperture loss corrections by dividing by encircled
energy fractions of 0.598 (F275W) and 0.611 (F814W) as prescribed in the data
analysis instructions from Space Telescope Science
Institute~\cite{Deustua2016,Bohlin2016}, then computed AB magnitudes using zero
points from the same publications.

\subsubsection{Lensing model and identification of multiple images}

We produced a gravitational lensing model of the source-lens system using the
software package \textsc{lenstool}~\cite{LENSTOOL}. 
Our lensing models of the lens-source system establish that the detected
occurrences of ionizing radiation in arc segments 1, 3 and 4 (Counterarc) are one
source, multiply imaged by the lensing potential.  
In fig.~\ref{fig:criticals}, we show our lensing model of Arc 1.  Three
foreground galaxies close to the line of sight of this arc produce local
perturbations to the global lensing potential, increasing the image multiplicity
compared to the case of a cluster lens without any such substructure.
Boundaries between separate (partial) images of the source galaxy are marked by
the critical lines where they cross the arc; we find 6 separate images  of the
Lyman continuum emitting clump in this arc segment. 

For arc segment 2, there are several plausible
models that are all consistent with the  
data, and which do not constrain the location of the critical curves
well. However, given the presence of only one LyC-emitting source in the other
arc segments, we find it most likely that the detections in this arc segment are
multiple images of the same image, too. 

This is supported by the MagE spectra described above.  Two of the spectroscopic
observations covered two of the images in Arc 2, with a third covering a
non-ionizing region in Arc 2 for comparison. In panels B--C in
figure~\ref{fig:magespecs}, we show plots of the spectra from the pointings
shown in figure~\ref{fig:magespecs}A, zoomed in on the stellar C~\textsc{iv}
1550~\AA\ and the Si~\textsc{iv} 1393,1402 \AA\ interstellar and circumgalactic
gas lines. The former is sensitive to age and individual characteristics of a
stellar population, the latter is sensitive to kinematics and ionization of the
surrounding gas. The spectra have been normalized by division by their median
value and smoothed by a 3 pixel boxcar kernel to reduce random noise, but have
not otherwise been manipulated.  The spectra from pointings with LyC detections
are nearly identical in both the stellar and the ISM line, while the
non-detection pointing is different in both. The spectra confirm the conclusion
that the LyC leaking regions are indeed multiple images of the same region,
derived independently from the lensing models discussed above.

\subsubsection{Stellar population synthesis\label{sec:sb99}}

The young, massive stars which produce Lyman continuum, have characteristic
spectral features in the rest-frame far-ultraviolet such as broad
N~\textsc{V}~1240~Å and C~\textsc{IV}~1550~Å stellar wind
profiles~\cite{Leitherer1999}, and weak photospheric absorption
lines~\cite{De-Mello2000}. These features constrain the age and metallicity of
the stellar population, and, consequently, the intrinsic ionizing continuum.

We constrained the ionizing continuum by fitting the observed, Milky Way
extinction-corrected MagE spectrum~\cite{Sunburst2017} with fully theoretical
stellar population models, following an established
methodology~\cite{Chisholm2015}. We used the spectral region between
1240--1900~\AA\ in the rest frame while masking regions of strong ISM absorption
and emission lines as well as absorption from intervening systems. We then
assumed that the far ultraviolet continuum is a discrete sum of multiple
single-aged populations of O- and B-type stars. Thus, we created a linear
combination of theoretical stellar templates, with ages varying between
1--40~Myr. Due to line-blanketing in the atmospheres of massive stars, the
stellar metallicity also sensitively determines the ionizing continuum and we
included stellar templates with metallicities of 0.05, 0.2, 0.4, 1.0, and 2.0
times the Solar metallicity~Z$_\odot$ to account for a wide range of possible
metallicities. The final suite of models consisted of 50 stellar templates (five
metallicities each with 10 possible ages) and we fit for a linear coefficient
multiplied to each individual theoretical stellar template using the IDL routine
\textsc{mpfit}\cite{MPFIT}.  The resulting linear combination of stellar models
was attenuated by a dust attenuation law appropriate for a $z\sim3$ starburst
galaxy~\cite{Reddy2016}.  The attenuation parameter was left as a free parameter
in our fits to find the value that best matched the observed continuum slope.
We found a best-fit extinction of $E(B-V) = 0.146 \pm 0.003$ mag., which yields
a light- and throughput-weighted attenuation in the two filters of
$A_{\mathrm{F275W}} = 2.06\pm0.11$, $A_{\mathrm{F814W}} = 0.92\pm0.03$.


We used the fully theoretical, high-resolution \textsc{Starburst99} stellar
continuum models, compiled using the \textsc{WM-Basic}
method~\cite{Leitherer2010} with the Geneva atmospheric models with high-mass
loss~\cite{Meynet1994}. We assumed a Kroupa initial mass function (IMF), with a
power-law index of 1.3 (2.3) for the low (high) mass slope, and a high-mass
cut-off at 100~M$_\odot$. The fitted stellar population is dominated by a very
young (a light-weighted age of 2.9~Myr), moderately metal-rich (0.56~Z$_\odot$)
stellar population. We tested whether the assumed \textsc{Starburst99}
theoretical stellar templates impacted the modeled ionizing continuum by fitting
the observations with \textsc{bpass}~\cite{Eldridge2017} models, which
produced 
similar ages, metallicities, and ionizing continua, largely because the two
libraries have similar O-type stellar models\cite{Eldridge2017}.

The high-resolution \textsc{Starburst99} models used for the fitting accurately
reproduce the narrow observed spectral features, but do not extend blueward of
900~\AA\ into the Lyman continuum~\cite{Leitherer2010}.  Using the linear
coefficients obtained from 
the high-resolution models, we created a low-resolution \textsc{Starburst99}
model, with and without attenuation. The extinction-free template models the
intrinsic ionizing continuum and allows us to compare the modeled and observed
Lyman continuum.

To investigate systematic uncertainties introduced by the choice of stellar
population and dust models, we also created a stellar population model based on
the \textsc{bpass} library~\cite{StanwayEldridge2018} in the same way, and
performed the same fits assuming an SMC-like dust attenuation
law~\cite{Gordon2003}. These are compared in the Supplementary text below.

\subsubsection{Milky Way dust correction}

All measured fluxes from \emph{HST} and MagE have been corrected for foreground
Milky Way dust using a reddening $E(B-V) = 0.09427$ mag~\cite{Schlegel1998}, and
assuming a standard extinction law with a standard $R_V=3.1$\cite{Cardelli1989}.
The effective wavelength for each of the \emph{HST} filters was determined as
the average wavelength in each filter, weighted by the products of the
uncorrected \textsc{Starburst99} model spectrum described above, and the
instrument throughput.  This procedure yields multiplicative dust correction
coefficients of 1.71 for F275W and 1.18 for F814W.

\subsubsection{Emission line diagnostics} 
\begin{figure}[h] 
	\centering
	\includegraphics[width=\columnwidth]{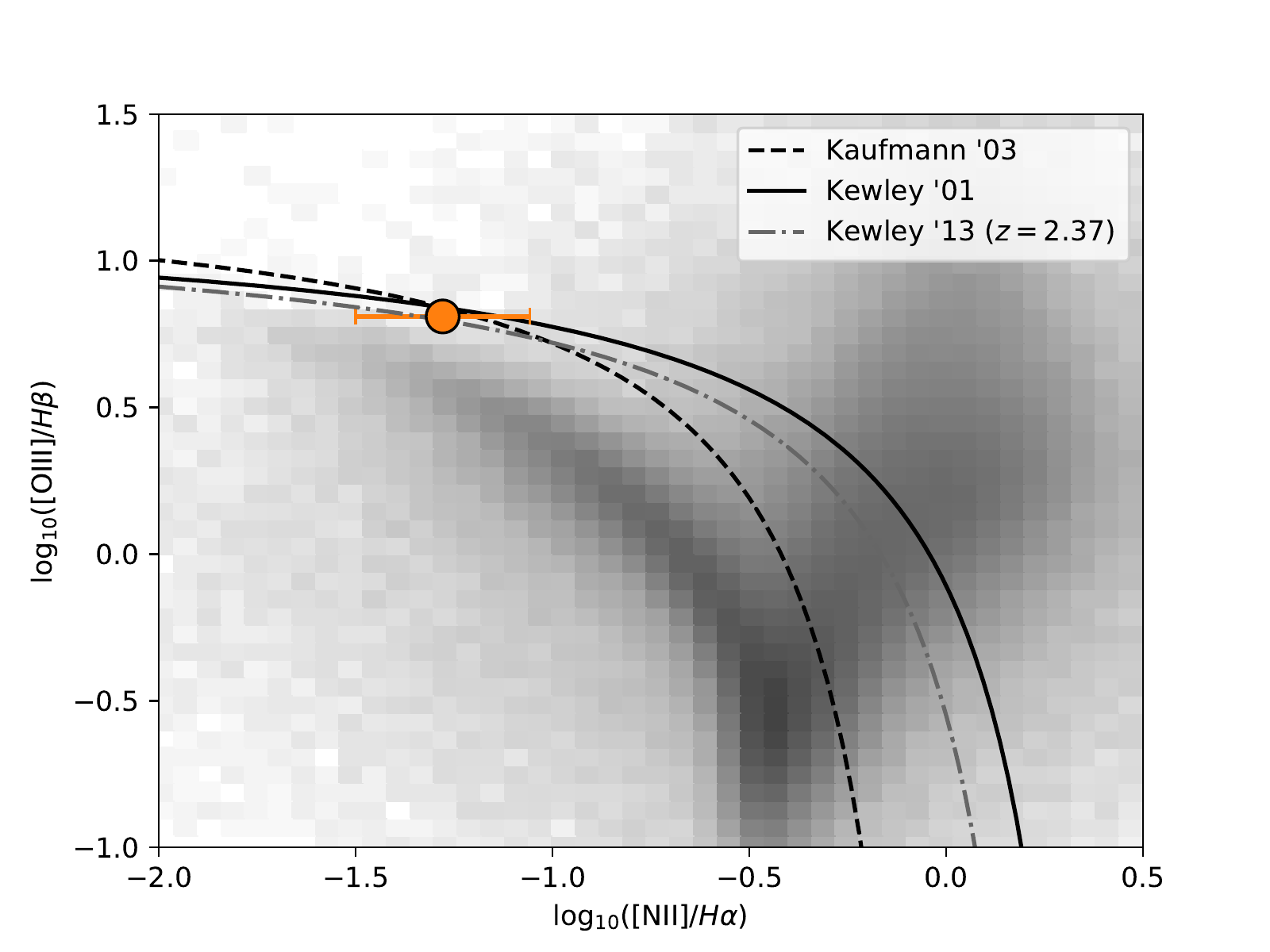}
	\caption[Baldwin, Phillips and Terlevich (BPT) diagram]{\textbf{Baldwin, Phillips and Terlevich (BPT) 
		diagram}~\cite{Baldwin1981} showing
		emission-line diagnostics of the Sunburst arc ionizing sources
		(orange point with error bars) based on Magellan/FIRE
		spectroscopy from previous work~\cite{Sunburst2017}. 
		Overlaid are theoretical and
		empirical stellar/AGN separation
		lines~\cite{Kewley2001,Kauffmann2003}.
		The grey-scale background shows a 2D histogram of 10.000 random objects from the
		Sloan Digital Sky Survey~\cite{SDSSIV}.  The grey dash-dotted 
		curve represents a fit to the main star formation locus at $
		z=2.4$~\cite{Kewley2013}.\label{fig:sunburstbpt} } 
\end{figure}

\begin{figure}[h] \centering
	\includegraphics[width=\columnwidth]{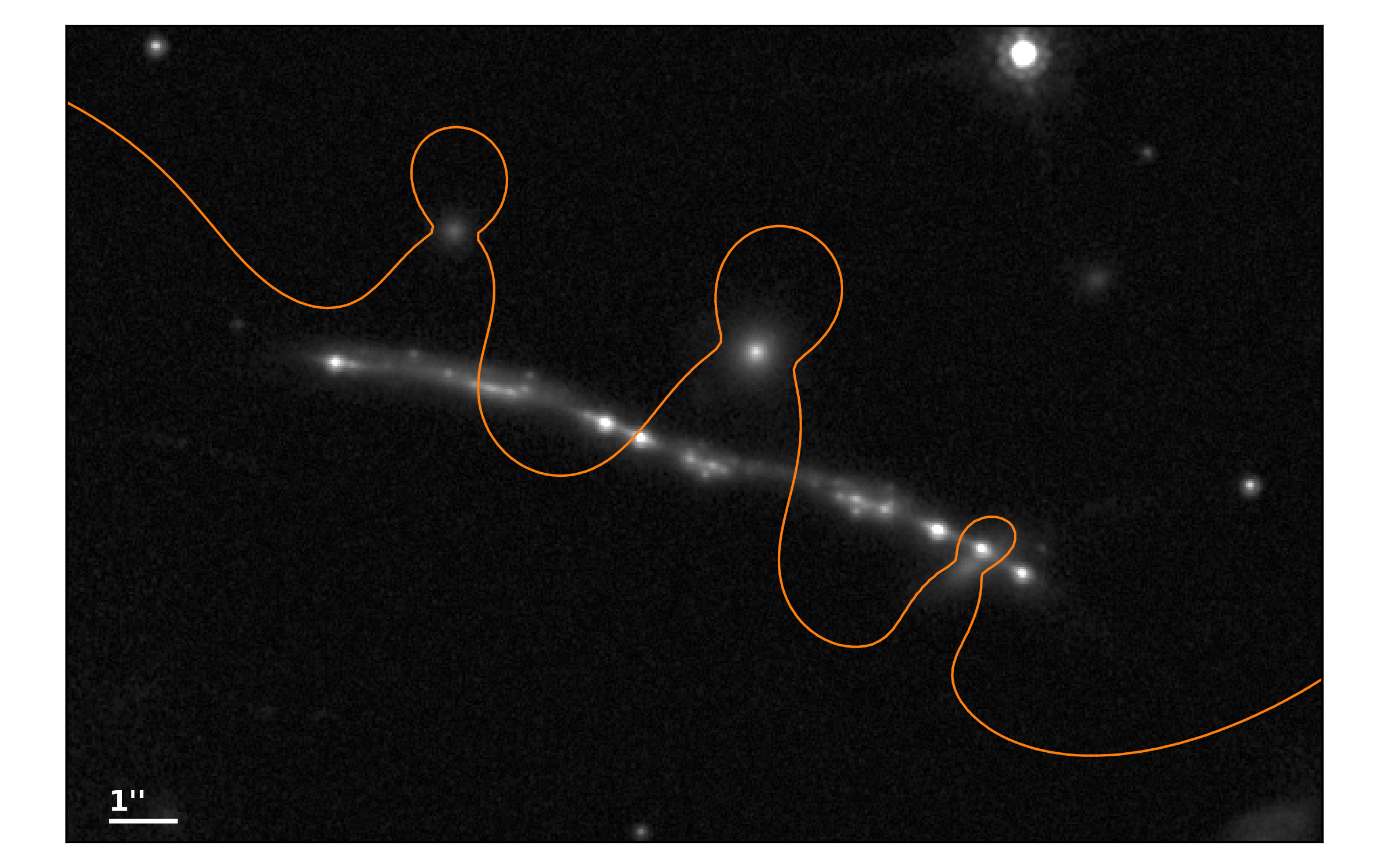}
	\caption[Critical lines in lensing model of Arc 1]{\textbf{Critical
		lines in the lensing model of Arc 1}. This figure shows Arc 1 as
		observed in F814W with the critical curve of the lens model
		overlaid in orange. Three members of the lensing
		cluster close to the line of sight produce perturbations to the
		large scale lensing potential which increase the image
		multiplicity compared to a smooth, non-perturbed potential.
		This arc segment contains 6 images of the leaking region (see
		main text). North is up, East is left. 
		These critical curves, along with the input
		file used to generate them, are available as a
		supplementary download (Data S1)\label{fig:criticals} 
	} 
\end{figure}

\begin{figure*}[!ht] \centering
	\includegraphics[width=.99\textwidth]{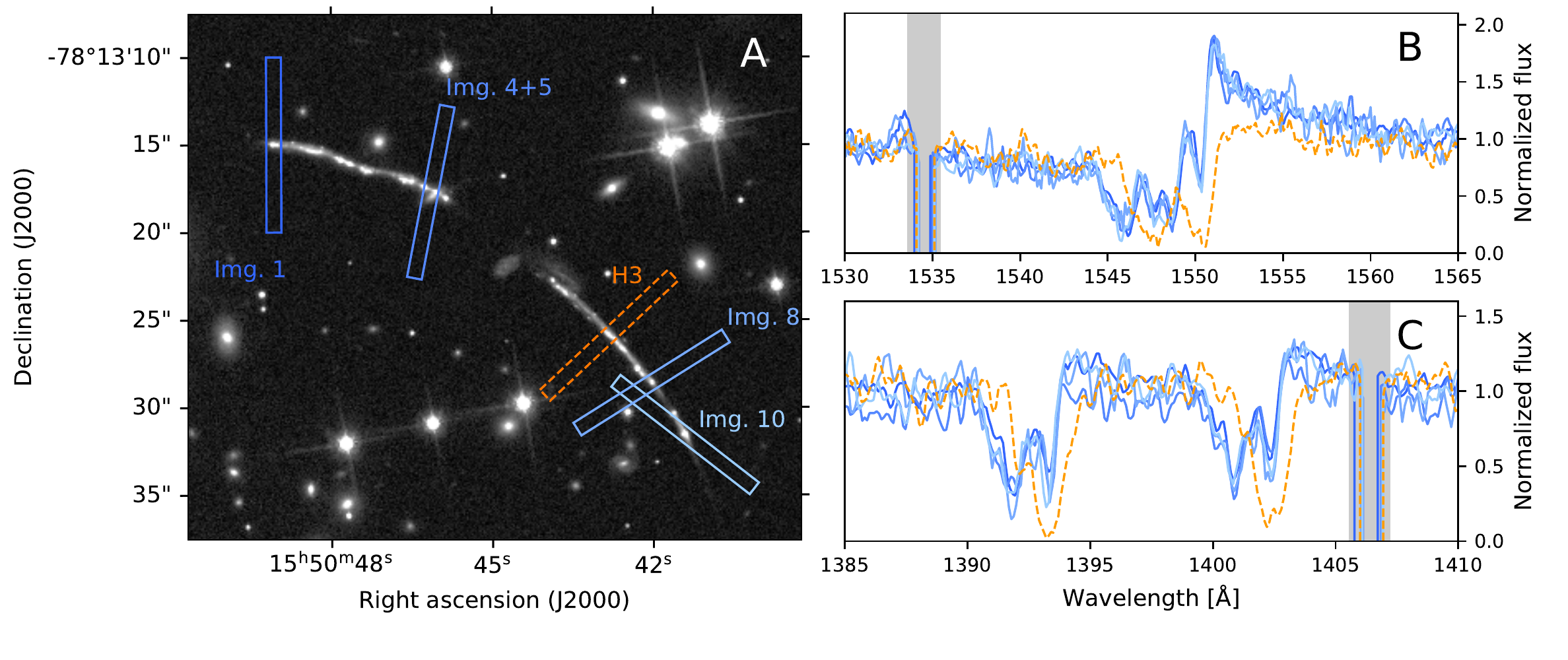}
	\caption[Spectroscopic comparison of leaking regions and non-leaking
	region]{\textbf{Spectroscopic comparison of leaking regions and
		non-leaking region.} Panel A shows the \emph{HST} ACS/F814W
		image zoomed in on the Arc1 and Arc2, containing images~1--10 of
		the LyC-leaking ionizing region, with locations of the MagE
		slits overlaid. Slits covering ionizing detections are blue and
		fully drawn, while the slit with no ionizing radiation detected
		is shown with orange dashes. The blue slits are named by the
		convention introduced in fig.~\ref{fig:cutouts}.  Panel B shows
		the median-normalized MagE spectra zoomed in on a narrow
		wavelength region containing the the C~\textsc{iv} 1550 Å
		stellar wind feature, and panel C the ISM/CGM
		Si\textsc{iv}~1393,1402 \AA\ doublet.  Spectra in panels B--C
		are colored and shown in line style corresponding to slit color
		and line style in panel A, and grey shading indicates data
		masked out due to detector gaps.\label{fig:magespecs} 
	} 
\end{figure*}

To test for signs of AGN activity, we
performed the line ratio based Baldwin, Phillips, and Terlevich
[BPT,~\cite{Baldwin1981}] diagnostic based on a rest-frame optical Magellan/FIRE
spectrum published in a previous work~\cite{Sunburst2017}, and compared it to
theoretical and empirical distinction criteria~\cite{Kauffmann2003,Kewley2006}
and a theoretical star formation locus accounting for the harder UV field at
$z\sim2.4$~\cite{Kewley2013}. The result is shown in
figure~\ref{fig:sunburstbpt}. We find no sign of AGN activity in the Sunburst
Arc.

\clearpage

\subsection{Supplementary text}

\subsubsection{Ionizing escape fractions\label{sec:fesc}}

The relative and absolute LyC escape fraction are defined as the fractions of
intrinsic photons that escape the gas (and dust) of the source galaxy and
reaches the IGM\@. We have computed this based on the synthetic
dust-absorbed and intrinsic spectra resulting from the stellar population
modelling described above. Focusing on the relative escape fraction, it
is defined as:

\begin{equation}
    f_{\mathrm{esc, rel}} =
    \frac{F_{\mathrm{275}}^{\mathrm{obs}}}{F_{275}^{\mathrm{int,ext}}}\label{eq:fescdef}
    \frac{1}{T_{\mathrm{IGM}}},
\end{equation}

where the numerator in the first fraction is the observed flux in the F275W
filter, and the denominator is the same as we would see it through a completely
ionized (but not dust-free) medium, and the transmission coefficient
$T_{\mathrm{IGM}}$ is the fraction of escaping LyC that passes through the IGM
without absorption. We do not know $F_{\mathrm{275}}^{\mathrm{int,ext}}$
directly, but since the non-ionizing continuum in F814W is unaffected by neutral
hydrogen, we can use the theoretical spectra to compute an expected flux in
F275W assuming complete transparency to LyC\@:

\begin{equation}
F_{\mathrm{275}}^{\mathrm{int,ext}} =
    \int L_{\mathrm{S99}} T_{\mathrm{275}} d\lambda
    \frac{F_{\mathrm{814}}^{\mathrm{obs}}}{\int L_{\mathrm{S99}}T_{\mathrm{814}}d\lambda}
    \frac{\int T_{\mathrm{814}}d\lambda}{\int T_{\mathrm{275}}d\lambda},
\end{equation}

where $F_{\mathrm{814}}^{\mathrm{obs}}$ is the observed flux in F814W,
$L_{\mathrm{S99}}(\lambda)$ is the theoretical spectral flux density from
\textsc{Starburst99}, and $T_{i}(\lambda)$ is the system transmission curves for
filter $i$. Combining this with eq.~\ref{eq:fescdef} and rearranging, we find:

\begin{equation}
f_{\mathrm{esc,rel}} T_{\mathrm{IGM}} = \frac{F_{\mathrm{275}}}{F_{\mathrm{814}}}
    \frac{\int T_{\mathrm{275}}d\lambda}{\int T_{\mathrm{814}}d\lambda}
    \frac{\int L_{\mathrm{S99}}T_{\mathrm{814}}d\lambda}{\int
    L_{\mathrm{S99}}T_{\mathrm{275}}d\lambda}.
\end{equation}

\noindent We find the apparent absolute escape fraction by the same procedure
for the unattenuated theoretical spectra from \textsc{Starburst99}; this yields
a multiplicative factor 

The escape fractions found this way are what we call the ``apparent escape
fractions'', as they do not account for absorption in the intergalactic medium.
These are shown in figure~\ref{fig:fesc} for each lensed image.

\subsubsection{Transmission in the intergalactic medium}

\begin{figure}[h] \centering
	\includegraphics[width=\columnwidth]{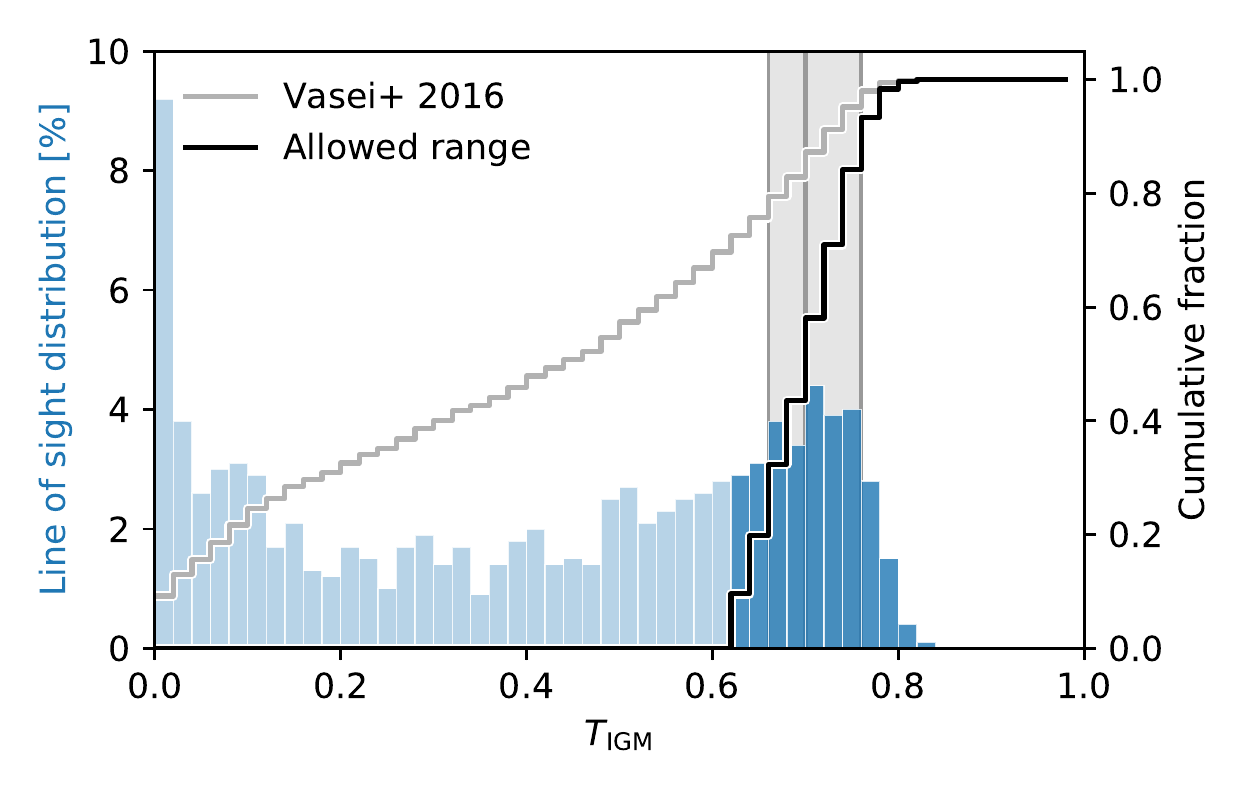}
	\caption[Simulated distribution of IGM transmission coefficients at
	$z=2.38$]{\textbf{Simulated distribution of IGM transmission
		coefficients at $z=2.38$.} IGM transmission histogram by Vasei
		et al.\cite{Vasei2016}. Values excluded because they lead to
		escape fractions above unity are shown in lighter shade.
		Grey steps show the original, and black steps the updated
		cumulated distribution derived from the remaining, permitted
		values of $T_{\mathrm{IGM}}$. The 16th, 50th and 84th percentile
		of the updated distribution are shown as vertical grey lines 
		with the region spanned by them shaded.\label{fig:vaseihist} 
	} 
\end{figure}

To estimate the IGM transmission, we have adopted an IGM transmission
distribution~\cite{Vasei2016}, which measures the IGM transmission out to
$z=2.38$ along simulated lines of sight.  This redshift is nearly identical to
that of the Sunburst Arc, so we adopt their coefficients without modifications.
The median coefficient $T_{\mathrm{IGM}}=0.4$ from that study
yields a relative escape fraction for the Sunburst Arc of more than 120\%. 
All coefficients $T_{\mathrm{IGM}} \lesssim 0.48$ are excluded in our case,
because they would yield escape fractions larger than 100\%.  With these
values excluded, we renormalized the remaining distribution and computed the
cumulative probability and found the median value  with 16 and 84\% confidence
levels.  The original and updated IGM transmission histograms, with cumulated
fractions, are shown in fig.~\ref{fig:vaseihist}.
%
The modified distribution yielded a median value with 16th and 84th percentile
confidence levels of $T_{\mathrm{IGM}} = 0.69^{+0.06}_{-0.04}$. 
For the measured apparent escape fraction
of image~12, this yields a relative escape fraction of $f_{\mathrm{esc, rel}} =
0.93^{+0.07}_{-0.11}$  and an absolute escape fraction of   $f_{\mathrm{esc,
abs}}=0.32^{+0.04}_{-0.06}$. We estimated the by approximating the
$T_{\mathrm{IGM}}$ distribution by an asymmetric Gaussian distribution,
interpreting the 16th and 84th percentiles as standard deviations, and
square-summing with the standard errors of the $f^*_{\mathrm{esc}}$ above.

\subsubsection{Transverse scale of IGM probed by sight lines to multiple images}

To calculate the transverse distances between sight lines, we used the
approximation of a spherically symmetric lensing system with the telescope
aligned with the source and the center of the lens. The ratio between transverse
distances in the lens plane and in any plane between the source and the lens is
then:

\begin{equation}
	\frac{d_i}{d_L} = \frac{\left[1 - \frac{D_{Li} D_{s}}{D_{Ls}
	D_{i}}\right]D_{i}}{D_L},\label{eq:lensdist}
\end{equation}

where $d$ is the transverse physical distance, $D = D(z)$ is the cosmological
angular diameter distance as a function of redshift, and the subscripts
$s$, $L$ and $i$ denote source, lens, and intervening plane.  In
fig.\ref{fig:sightlines}, we plot the transverse, physical distances
corresponding to $1'', 10'' \textrm{ and } 55''$ in the lens plane, as function
of redshift and co-moving distance. These angles are the approximate distances
between images~2 and 3, across Arc 1 between images~1 and 6, and across the
entire arc between images~1 and 12.

Because we found no significant contribution from differential magnification to
the differences in apparent escape fraction between images in the arc, we
conclude that they arise from changing column densities of neutral hydrogen
along the lines of sight. Photons of wavelength longer than the \Lya\ line at
$\lambda = 1216$ Å are unaffected by neutral hydrogen, so absorption variations
must occur before cosmic expansion has redshifted all the intrinsically ionizing
photons beyond this wavelength. In addition, it is unlikely that the variations
occur due to \Lya\ absorption in the IGM:\@ Spectral \Lya\ absorption features
are typically narrow and strongly saturated, and a change in column density of a
factor of a few will have only a small effect on the total absorbed flux.  A
difference of a factor 2 in transmitted flux due to varying \lya\ absorption
would require a doubling in the number of absorbing systems.  In contrast, when
the photons are still within the ionizing wavelength range, the transmission
depends directly on the logarithm of the total column density integrated over
the relevant wavelength range, and the same variation in absorption would only
require a modest change in integrated H\textsc{i} column density. This leads us
to conclude that the absorption most likely happens when the light still has
ionizing wavelengths.

As the lower redshift limit, we have adopted a redshift of absorption of
$z_{\mathrm{abs}}\gtrsim2.0$, at which half of the flux observed in F275W has
redshifted out of the ionizing range. Because we have found variations in the
apparent escape fraction of almost a factor of 5, this is a conservative
estimate. At this redshift, the transverse distance between the lines of sight
towards images~2 and 3 is determined by eqn.~\ref{eq:lensdist} to be
$d_{23}=331$~pc.  The upper redshift limit depends strongly on the physical
scale of the region emitting ionizing photons. In the upper limit, the effective
diameter of the knot is $\sim 160$ pc. In order to produce a difference in
$f^*_{\textnormal{esc}}$ of a factor 2, as was the adopted requirement above,
this means the transverse distance must be at least half the effective diameter,
or around 80 pc.\ which, using eqn.~\ref{eq:lensdist}, yields
$z_{\textnormal{abs}} \lesssim 2.27$. However, the cluster may be considerably
smaller than this; studies of the local Universe have found massive, luminous
starburst regions as small as $\sim 5$ pc.\cite{Adamo2013}, and the ionizing
emission may originate from an even more compact stellar association. If this is
the case, the absorption could take place in the circumgalactic medium (CGM) of the
galaxy. In the extreme limit, all the ionizing radiation may originate from one
or two extremely young, bright, massive stars, in which case the absorption
could in principle take place in the interstellar medium of the galaxy itself,
on scales of less than a parsec.

We speculate that the most likely interpretation of these variations is that
they probe variation on the scale of a few tens or hundreds of parsec, either in
a Lyman-limit system withing $\sim 500$ Mpc.~from the galaxy, or on a scale of
$\lesssim 1$ parsec in the galaxy's own CGM.\@
\subsubsection{Differential magnification}

One possible explanation for the variation in the F275W/F814W flux ratios
between the lensed images of the leaking region is differential magnification:
If the sources of emission in F275W and F814W are not completely coincident (if
e.g.\ the ionizing radiation is dominated by one massive Wolf-Rayet star
displaced from the central stellar component), the sources and the lens caustics
might be arranged in such a way as to magnify one component stronger than the
other. However, this is mainly a concern when the caustics are actually
crossing, or very close to, the bright sources, which makes it unlikely that
this effect dominates the variations we observe. The distance between the
components, if any, is unresolved in our observations and thus known to be much
smaller than the distance from either to the critical lines. Still, to test this
further, we consider the following:

Because the caustics do not cross the emitting region, differential
magnification may only occur if one component is closer to the caustics than the
other. If the center of flux in F814W is closer to the caustic than that of
F275W, the stronger magnification of the non-ionizing flux will yield a lower
apparent escape fraction, and vice versa.

This effect is somewhat counteracted by the presence of an extended stellar
component surrounding the central, unresolved peak in F814W.  In the case where
the F275W source is more strongly magnified, a larger contribution from this
extended component will be present in the aperture in F814W, but absent in
F275W, and vice versa.  This will counteract the effect described above.
However, since gravitational lensing preserves surface brightness, the
contribution from the extended component will change more slowly
than the main source.  Thus, despite the presence of this effect, we still
expect to see a strong correlation between the measured F814W flux (which is
unaffected by neutral hydrogen absorption) and derived apparent escape fraction,
if the effect is due to differential magnification.

\begin{figure}[h]
	\centering
	\includegraphics[width=\columnwidth]{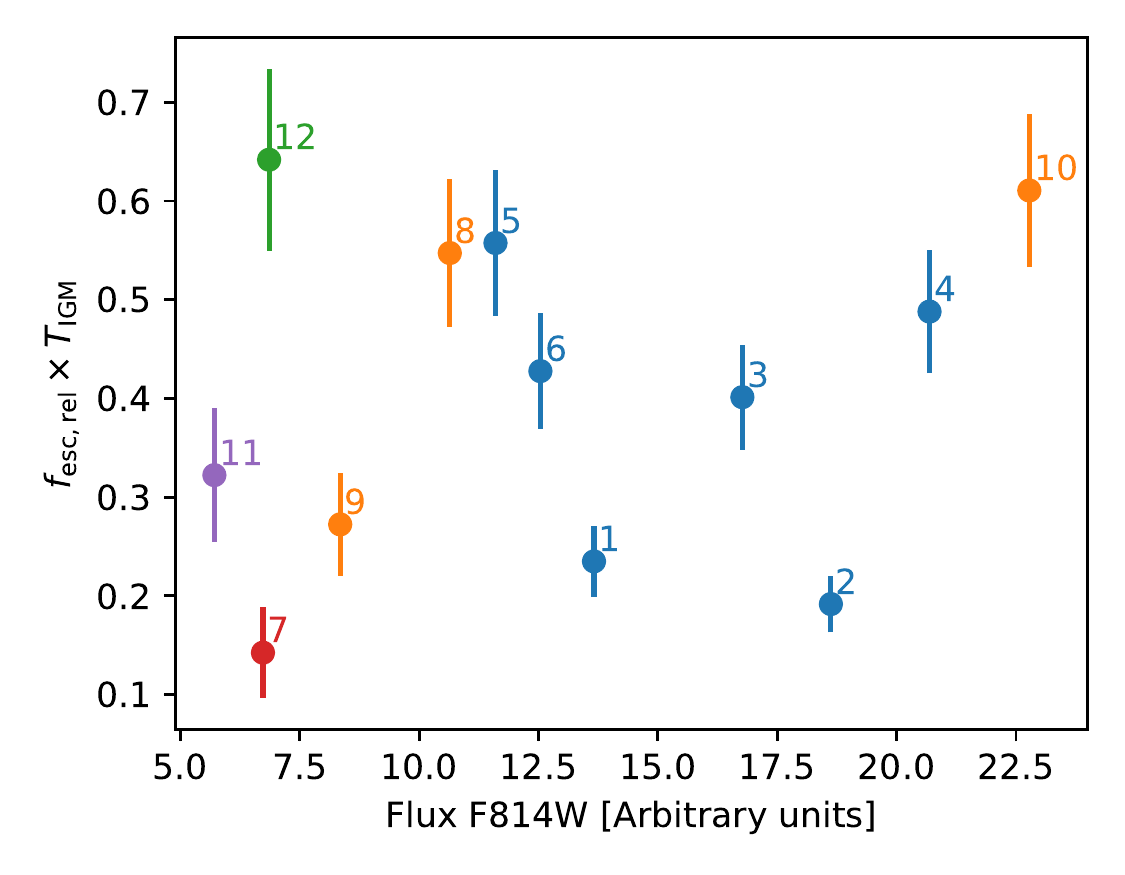}
	\caption[Flux in F814W vs. apparent escape fraction.]{\textbf{Flux in 
		F814W vs. apparent escape fraction.} $F_{\mathrm{814}}$ vs.\ 
		apparent escape fraction is shown for each of the 12 images,
		with colors following the scheme established in
		figure~\ref{fig:cutouts} and ~\ref{fig:fesc}\label{fig:diffmag}
	}
\end{figure}

In fig.~\ref{fig:diffmag}, we show a plot of the F814W fluxes vs.\ the apparent
escape fractions.  We find no significant correlation, with a measured Pearson's
$r = 0.2$ and a p-value of $p = 0.53$, leading us to conclude that
this effect is not what causes the found variations.

\subsubsection{Systematic uncertainties}

\noindent \textbf{Non-ionizing contamination in F275W}\\
A small, but non-negligible amount of the light in F275W is transmitted redward
of the observed wavelength of the Lyman edge. To ensure we are not just
observing non-ionizing continuum, we have computed the expected flux in the
filter by multiplying the synthetic \textsc{Starburst99} spectrum by the
transmission curve of F275W and integrating this on the red side of the Lyman
edge only. The derived fluxes, which span from $\sim2\%$ to $\sim10\%$ of the
measured fluxes, were then subtracted from the measured F275W fluxes to correct
for the contamination.  All properties derived from measured F275W are corrected
for this effect.\\

\noindent \textbf{Photometric aperture does not match spectroscopic slit}\\
The spectroscopic slit, on which our stellar population modeling is based, is
considerably larger than the photometric apertures. The line-of-sight ionizing
escape fraction reported in this work is based on the assumption that the UV
spectroscopic signature of the stellar population in the photometric aperture is
similar in composition to that of the total population inside the spectroscopic
slit. If this is not the case, we may have estimated the intrinsic LyC flux in
the photometric aperture incorrectly, and consequently reached an incorrect
escape fraction.
If 
the photometric aperture preferentially contains stars at the 
top end of the LyC luminosity function, the intrinsic LyC flux could be higher
than what is predicted from averaging over the whole spectrographic slit.
Conversely, if the aperture is 
populated with the faintest LyC sources, our computed escape fraction is too
low.
 
To estimate how strong these effects are, we have smoothed the F814W exposure of
the arc to a seeing of 0\farcs75 to match that of the MagE observations. We then
modeled the smoothed flux inside the MagE slit by three spatial components: A
bright, narrow, unresolved profile coincident with the LyC detection and the
photometric aperture, a slightly broader Sersic~\cite{Sersic1963} profile at the
same location (consistent with hot, star-forming regions like e.g. 30 Doradus in
the Large Magellanic Cloud), and a third, extended, less regular component. We
found that about 40\% of the light in the slit was concentrated in the central,
unresolved spike. 

Stellar population modeling determines the contribution by stellar light
fraction to the LyC flux; we can thus compute how much higher or lower the
intrinsic LyC would be if these 40\% in the central spike come exclusively from
the higher or the lower end of the ionizing-to-nonionizing flux ratio
distribution. We find that in the extreme case, $f_{\textnormal{esc, rel}}$
could be as high as 93\% and as low as 62\%.  This is however an unlikely,
extreme scenario. The stellar population synthesis model implies that close to
100\% of the rest-frame far-ultraviolet emission observed in F814W is generated
by stars with ages less than 4 Myr, typical of a single burst episode, in which
stars of various ages are most likely distributed more evenly, and the
uncertainties from this effect thus considerably smaller in the prese case,
possibly even negligible.\\

\noindent \textbf{Mismatch of wavelength ranges for photometry and stellar
	population modeling}\\
The stellar population models are based on the wavelength range between 1240 and
1900 \AA, while the non-ionizing photometry used to calibrate this model and
predict the intrinsic flux in F275W is based on the F814W band, in which we are
not able to test whether the spectrum of the galaxy still matches the model
closely. Any such mismatch would lead to an over- or underestimation of the
intrinsic flux in F275W based on the observed flux in F814W, which in turn leads
to an erroneous estimate of the LyC escape fraction.

The light weighted stellar population age arrived at above is very low
($\sim$2.9 Megayears), and the slope of the UV spectrum decreases with age,
meaning that any such mismatch would likely be in the shape of an
over-estimation of the intrinsic F275W flux relative to the F814W flux, and thus
an under-estimation of the escape fraction inferred from observed F275W flux.
However, given the high escape fraction found, this effect is unlikely to be
strong.\\

\begin{figure*}[h]
	\centering
	\includegraphics[width=\textwidth]{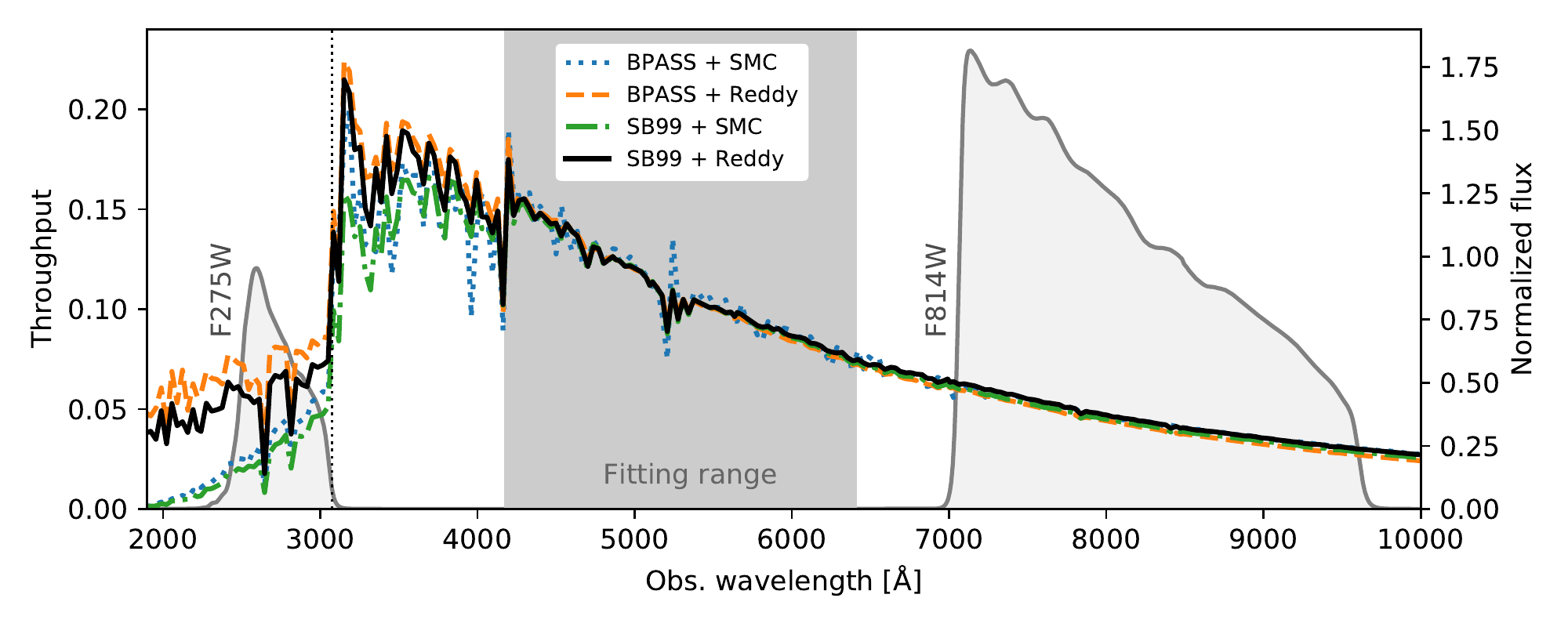}
	\caption[Comparison of stellar population models]{\textbf{Comparison of
	stellar population models:} Comparison of normalized 
	\textsc{Starburst99} and \textsc{bpass} models with the Reddy 2016 and 
	SMC dust extinction laws (see Materials and Methods above). The 
	wavelength range used for stellar synthesis fitting is shaded in grey. 
	Also shown are the F725W and F814W throughput curves. The reference
	model used in this work is the Starburst99 + Reddy 2016 model (black
	line).
\label{fig:starmods}}
\end{figure*}

\noindent \textbf{Stellar population models}\\ 
The escape fraction derived from the measured F275W/F814W flux ratios rest on
the assumed stellar population models. We have built these on the
\textsc{Starburst99} library, which is a well established and thoroughly tested
library of stellar population models for galaxies and star-forming regions. To
test how strong the assumptions built in to the stellar models affect the
derived escape fractions, we constructed an alternative model based on the
\textsc{bpass} library~\cite{StanwayEldridge2018}. The best-fit \textsc{bpass}
model~\cite{Chisholm2019} has age, metallicity and non-ionizing UV spectral
shape consistent with the \textsc{Starburst99} model, but predicts a higher
intrinsic LyC flux (see figure~\ref{fig:starmods}). We infer a lower relative
and absolute escape fraction than from \textsc{Starburst99}, with
$f_{\mathrm{esc,rel}}^{\mathrm{BPASS}} = 76^{+17}_{-8}\%$, and
$f_{\mathrm{esc,abs}}^{\mathrm{BPASS}} = 24^{+5}_{-3}\%$.  \\

\noindent \textbf{Dust attenuation model}\\
To model the stellar population from the spectra, we have made assumptions about
the dependence on wavelength of dust attenuation in the source galaxy. We have
assumed an average attenuation model calibrated towards starburst galaxies at
$z~3$~\cite{Reddy2016}, appropriate for the Sunburst Arc, and which is
observationally calibrated far into the ultraviolet waveleng range. However,
given the high magnification and small physical size of the regions studied in
this work, it is possible that a single line-of-sight extinction law like that
of the the SMC law~\cite{Pei1992}. This law is steeper in the far-UV than the
Reddy law, and predicts a lower intrinsic LyC flux and thus a higher inferred
escape fraction for a given observed flux (see figure~\ref{fig:starmods}). We applied the SMC law to both
\textsc{Starburst99} and \textsc{bpass} models and found that for the former,
the SMC law yielded a relative escape fraction of $\sim180\%$ even when assuming
a completely transparent IGM, while the latter yielded an escape fraction of
100\% under the assumption of a fully transparent IGM.\@ We conclude that the
application of the SMC extinction law to the Sunburst Arc yields unphysical
results.
The Reddy attenuation law is flat in the far-UV compared to other standard dust
extinction models. Generally, a steeper law leads to fainter predicted intrinsic
LyC and thus a higher inferred escape fraction for a given observed LyC flux.
Because the assumption of the Reddy law with the \textsc{Starburst99} models
yields an escape fraction as high as 93\%, we conclude that the dust attenuation
curve for the Sunburst Arc is indeed likely to be well approximated by the Reddy
law. \\

\clearpage

\subsection{Supplementary figures}

\end{document}